# An Information-Theoretic Method for Dynamic System Identification With Output-Only Damping Estimation

**Marios Impraimakis**[1]
Department of Mechanical Engineering,
University of Bath,
Bath BA2 7AY, UK
e-mail: mi595@bath.ac.uk

**Feiyu Zhou**
Department of Mechanical Engineering,
University of Bath,
Bath BA2 7AY, UK
Department of Civil Engineering,
Zhejiang University, Hangzhou 310027, China

**Andrew Plummer**
Department of Mechanical Engineering,
University of Bath,
Bath BA2 7AY, UK

*The system identification capabilities of a novel information-theoretic method are examined here. Specifically, this work uses information-theoretic metrics and vibration-based measurements to enhance damping estimation accuracy in mechanical systems. The method refers to a key limitation in system identification, signal processing, monitoring, and alert systems. These systems integrate various components, including sensors, data acquisition devices, and alert mechanisms. They are designed to operate in an environment to calculate key parameters such as peak accelerations and duration of high acceleration values. The current operational modal identification methods, though, suffer from limitations related to obtaining poor damping estimates due to their empirical nature. This has a significant impact on alert warning systems. This occurs when their duration is misestimated; specifically, when using the vibration amplitudes as an indicator of danger alerts for monitoring systems in damage or anomaly detection scenarios. To this end, approaches based on the Shannon entropy and the Kullback–Leibler divergence concept are proposed. The primary objective is to monitor the vibration levels in near real-time and provide immediate alerts when predefined thresholds are exceeded. In considering the proposed approach, both new real-world data from the multi-axis simulation table at the University of Bath, as well as the benchmark International Association for Structural Control–American Society of Civil Engineers (IASC–ASCE) structural health monitoring problem are considered. Importantly, the approach is shown to select the optimal model, which accurately captures the correct alert duration, providing a powerful tool for system identification and monitoring.* [DOI: 10.1115/1.4071371]

## 1 Introduction

Early warning system monitoring and identification are critical aspect for maintaining the safety against the effects of natural and man-made hazards. One of the main challenges is related to ensuring that systems and structures are represented by accurate updated models and digital twins [1–3], and by sensitive damage classification indicators [4,5]. A model updating representation is examined here using a novel information-theoretic approach. The information theory includes both the probabilistic and the non-probabilistic approaches [6]. It refers to the measurement of uncertainty in a set of data [7]. Nonprobabilistic approaches lie in the area of physics and thermodynamics, where entropy is used to quantify the disorder of a physical system.

The Shannon entropy is a central concept used in information theory to measure the uncertainty of a random variable. To this area, Papadimitriou et al. [8] proposed an approach to optimally locate sensors in a structure based on information entropy. Additionally, Yuen et al. [9] developed a sensor placement methodology for scenarios with uncertain excitations, such as earthquakes [10–12] and wind [13]. Furthermore, Papadimitriou [14,15] provided an asymptotic approximation for information entropy and extended the concept to multiple model classes. In other applications, Yan and Gao [16] derived an approximate entropy for damage detection, a statistical measure that quantifies the regularity of a time series. Interestingly, Green et al. [17] approximated the effect of a training dataset on parameter uncertainty and the plausibility of candidate model structures using information theory. In the same direction, Garcia Gonzalez et al. [18] developed a method based on Shannon Entropy for feature extraction of optical patterns, enhancing the accuracy of optoelectronic signal classifiers in structural health monitoring. Other measures for structural health monitoring include the joint entropy [19], the mutual information [20], the transfer entropy [21,22], the permutation entropy [23,24], the wavelet entropy [25], and the spectral entropy [26].

Along these lines, the Kullback–Leibler divergence quantifies how much one probability distribution differs from another. In this area, Tian et al. [27] utilized the Kullback–Leibler divergence for damage detection, introducing two additional divergences and a statistical distribution to mitigate asymmetry issues. Xie et al. [28] proposed detecting incipient damage in complex dynamic systems using the Kullback–Leibler divergence, where subspace identification [29,30] identifies dynamic models, and the Kullback–Leibler divergence examines changes in probability density functions between a reference dataset and online measurements. In another application, Tsioulou and Galasso [31] applied the Kullback–Leibler divergence to measure the similarity of probability distributions between recorded and simulated ground-motion intensity measures. Kullback–Leibler divergence is also an important concept in machine learning for structural health monitoring and engineering applications [32,33] to measure how one probability distribution differs from a second, usually a true or target distribution. If the Kullback–Leibler divergence is low, it indicates that a machine learning model captures well the patterns in the data. It is crucial in various machine learning tasks such as classification, clustering, and generative modeling [34]. More specialized tools are focusing on the shape of the uncertainty distribution or other special characteristics such as the Renyi's entropy [35], the Jensen–Rényi divergence [36], the Jensen–Shannon divergence [37], the f-divergence [38], the Tsallis entropy [39], the sample entropy [40], the Fuzzy entropy [41], the Fisher information [42], and the cross-entropy optimization [43].

---

[1]Corresponding author.
Manuscript received December 16, 2025; final manuscript received March 3, 2026; published online March 31, 2026. Assoc. Editor: Shengxi Zhou.

Relating to other applications of information entropy, information gain is also a concept used in the field of machine learning and in decision trees to determine the best attribute to split the data. Here, Chen et al. [44] introduced an information-based crack detection method to robustly characterize and detect cracks. Bertola et al. [45] proposed several definitions and metrics to quantify the information gained from monitoring data to better evaluate the benefits of monitoring techniques applied to a full-scale bridge case study in Switzerland. The value of information, on the other side, measures the benefit gained from obtaining and utilizing information to make decisions or take actions. In decision-making [46–49], having relevant and accurate information can lead to better decisions. Straub [50] discussed its calculation when data is gathered to enhance the reliability of engineering systems, illustrating how structural reliability techniques can effectively model the value of information. Later, Zonta et al. [51] created a logical framework to assess the impact of structural health monitoring on decision-making, demonstrated through a case study of the Streicker Bridge. In a specific application, Sykora et al. [52] provided practical guidance for the value of information estimation within an objective framework, quantifying the value of structural health monitoring in an engineering context applied to masonry structures.

However, current model updating approaches do not provide a framework for accurately modeling alert-type excitations. The issue is derived when poor prediction of vibration damping [53] leads to inaccurate early warning systems. Specifically, alert systems are highly sensitive to the damping behavior as it defines the duration of a serious event [54]. This poses a significant challenge for current approaches to be used for near real-time systems, especially when sensitive noncontact approaches are examined [55,56]. This work uses a novel information-theoretic approach to address this issue, enabling more accurate model-updating procedures for these systems.

The paper is organized as follows: Sec. 2 presents both the descriptive and mathematical framework for vibration monitoring, including the components and the methodology used. Section 3 provides an investigation of new real-world data from the multi-axis simulation table at the University of Bath. Section 4 applies the methodology to the IASC–ASCE (International Association for Structural Control–American Society of Civil Engineers) benchmark structural health monitoring problem. Section 5 provides a damage and anomaly detection application of the methodology. Section 6 discusses the results and gives potential future directions for research, while Sec. 7 draws the conclusions.

## 2 The Information-Theoretic Method Based on the Kullback–Leibler Divergence and the Signal Energy

A typical early warning system integrates near real-time monitoring and an internet-based alert system to enhance system monitoring. The system begins with the deployment of distributed sensors that continuously capture vibrations, displacement, and strain. These sensors, after processing on a local computer, communicate with a cloud-based or edge-processing system, which assesses structural integrity [57,58]. Real-time visualizations display time-series data allowing decision-makers to assess the situation and take appropriate action. If a critical threshold is exceeded, automated shutdown protocols for machinery or temporary structural reinforcements may be triggered. The system can also be extended to integrate predictive maintenance capabilities, using long-term monitoring data to forecast potential structural issues before they reach dangerous levels. The dynamic behavior of the system is represented by

$$\mathbf{M}\ddot{\mathbf{x}}(t) + \mathbf{C}\dot{\mathbf{x}}(t) + \mathbf{K}\mathbf{x}(t) = \mathbf{f}(t) \qquad (1)$$

where $\mathbf{M}$, $\mathbf{C}$, and $\mathbf{K}$ are the mass, damping, and stiffness matrices, respectively, $\mathbf{x}(t)$ is the displacement vector, and $\mathbf{f}(t)$ is the unknown stochastic-type excitation. Only the output $\mathbf{y}(t)$ is usually measured in large-scale infrastructure systems under operation conditions, which is expressed as

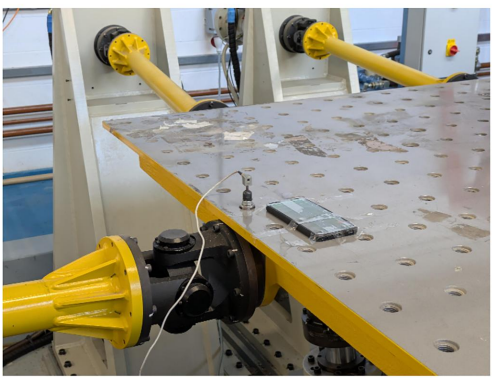
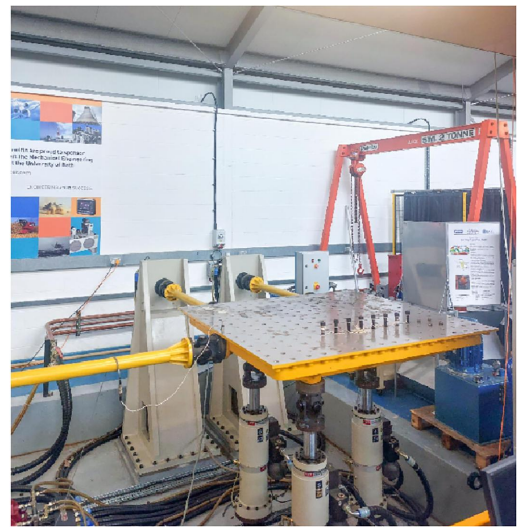

Fig. 1 The multi-axis simulation table at the University of Bath with the reference accelerometer for the Sec. 3 application (Photo by Jens Roesner)

$$\mathbf{y}(t) = \mathbf{C}_d \mathbf{x}(t) \tag{2}$$

where $\mathbf{C}_d$ is the output matrix. In the frequency domain, the power spectral density of the output is given by

$$\mathbf{S}_{yy}(\omega) = \mathbf{H}(\omega)\mathbf{S}_{ff}(\omega)\mathbf{H}^H(\omega) \tag{3}$$

where $\mathbf{H}(\omega)$ is the frequency response function matrix, $\mathbf{H}^H(\omega)$ is the Hermitian transpose of $\mathbf{H}(\omega)$, and $\mathbf{S}_{ff}(\omega)$ is the power spectral density of the stochastic input [59]. Since the input force $\mathbf{f}(t)$ is unknown in output-only modal analysis, the input is assumed to be a broadband stochastic excitation, namely, $\mathbf{S}_{ff}(\omega)$ is approximately constant over frequency. Here, $\mathbf{S}_{yy}(\omega)$ is computed directly from the output for output-only implementation instead of the previous equation. To extract modal parameters, singular value decomposition is applied to the spectral density matrix

$$\mathbf{S}_{yy}(\omega) = \mathbf{U}(\omega)\mathbf{\Sigma}(\omega)\mathbf{U}^H(\omega) \tag{4}$$

where $\mathbf{U}(\omega)$ contains the singular vectors representing the mode shapes, and $\mathbf{\Sigma}(\omega)$ is a diagonal matrix containing the singular values indicating the strength of each mode. Here, by identifying peaks in the first singular value curve, the system natural frequencies are determined. The energy of a signal is then used to examine its total magnitude over time or frequency. For a continuous-time signal $x(t)$, the signal energy $E$ is defined as

$$E = \int_{-\infty}^{\infty} |x(t)|^2 dt \quad (5)$$

while, for a discrete-time signal $x[k]$, the energy is expressed as

$$E = \sum_{k=-\infty}^{\infty} |x[k]|^2 \quad (6)$$

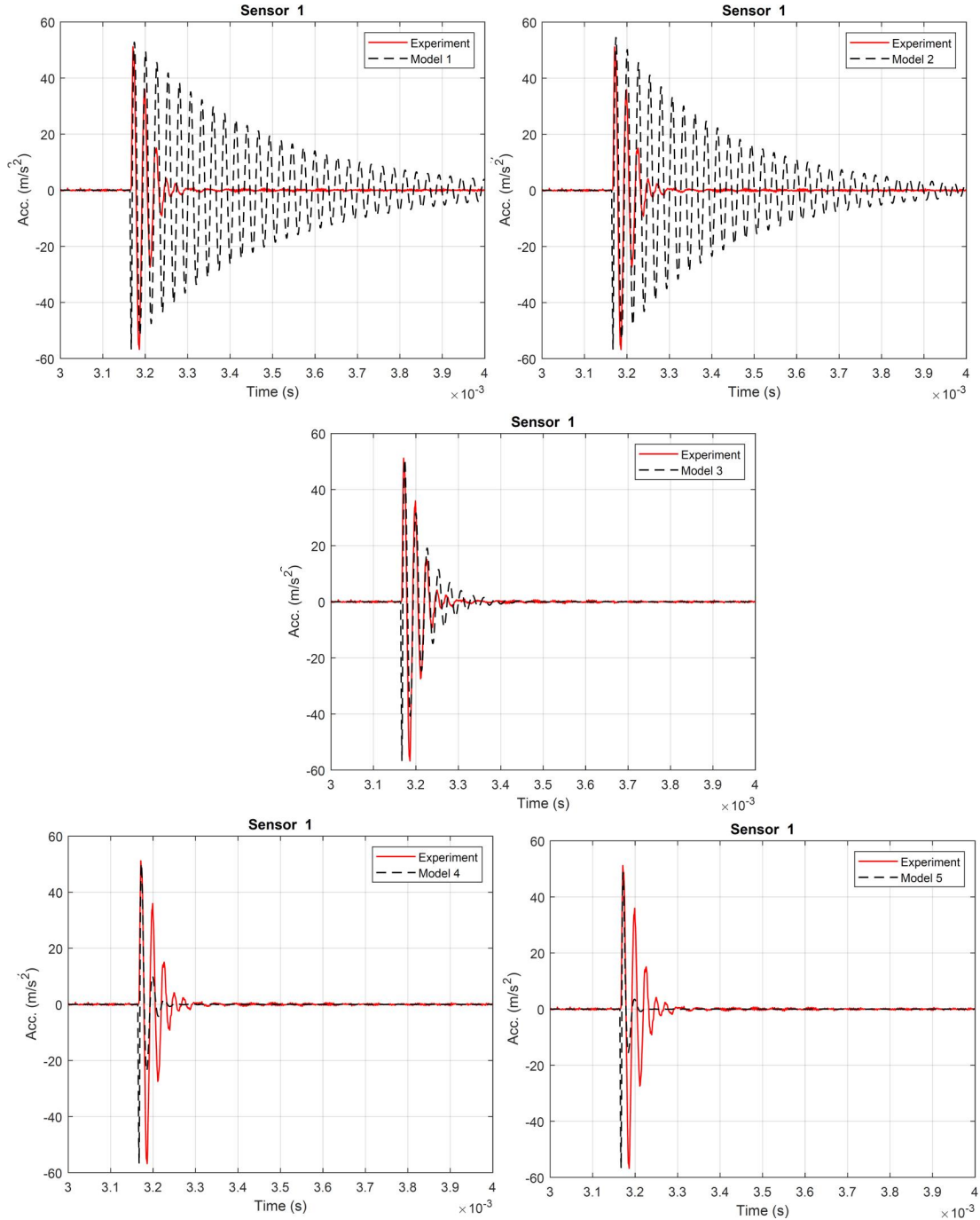

Fig. 2  Time-history performance of the Sec. 3 model for all damping models 1 to 5

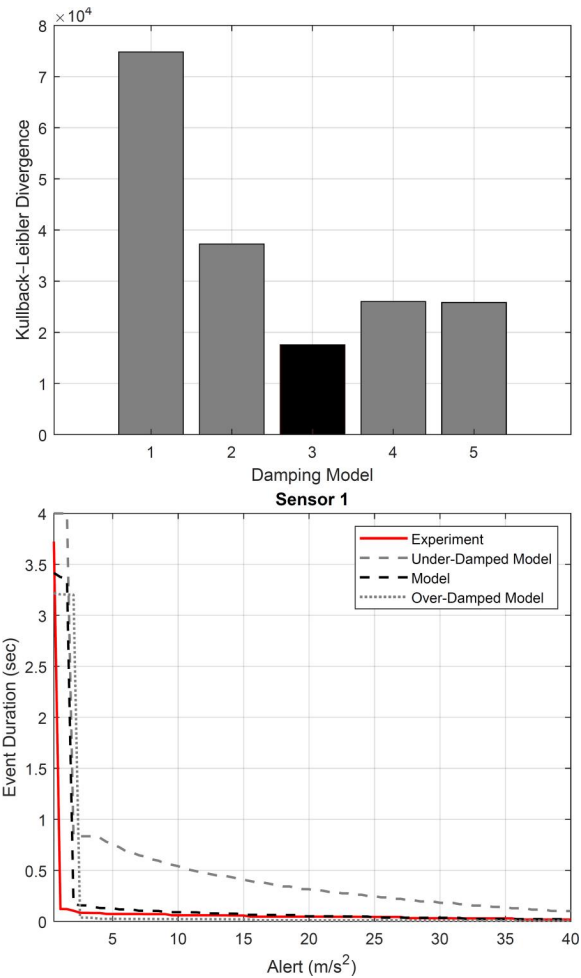

**Fig. 3 The Kullback–Leibler divergence of the Sec. 3 system, and the event duration between the experimental data and the model for various alert thresholds for three damping models. This plot shows the optimal model, as well as the under-damped model 1 and the over-damped model 5.**

Signal energy represents the power of a signal integrated over time or frequency. In structural health monitoring, the signal energy is used to evaluate the strength of vibration responses and monitor changes in structural behavior. In this work, the signal energy of the residual of the model $x[k]$ and the output measurement $y[k]$ is estimated as

$$E = \sum_{k=-\infty}^{\infty} |x[k] - y[k]|^2 \quad (7)$$

For a discrete random variable $Z$ with probabilities $p(z)$, the Shannon entropy $H(Z)$ is defined as

$$H(Z) = -\sum_{z \in Z} p(z) \log p(z) \quad (8)$$

where $Z$ is the set of possible values of $z$, and $p(z)$ is the probability of $z$. For continuous random variables, the entropy is generalized to the differential entropy

$$H(Z) = -\int_{-\infty}^{\infty} p(z) \log p(z) dz \quad (9)$$

where $p(z)$ is the probability density function. The Kullback–Leibler is mathematically expressed with a true distribution $P$ and an approximate or assumed distribution $Q$. For discrete distributions, the Kullback–Leibler divergence $D_{KL}(P||Q)$ is defined as

$$D_{KL}(P||Q) = \sum_{z \in Z} p(z) \log \frac{p(z)}{q(z)} \quad (10)$$

For continuous distributions, the Kullback–Leibler divergence becomes

**Table 1 Identified frequencies, damping ratios, and the mode shapes for the Sec. 4 system**

| Mode | 1 | 2 | 3 | 4 | 5 | 6 | 7 | 8 | 9 | 10 |
|---|---|---|---|---|---|---|---|---|---|---|
| f (Hz) | 2.564 | 4.150 | 8.362 | 13.611 | 16.052 | 23.315 | 25.269 | 32.96 | 39.063 | 43.152 |
| $\zeta$ | 0.013 | 0.013 | 0.013 | 0.013 | 0.013 | 0.013 | 0.013 | 0.013 | 0.013 | 0.013 |
| $\Phi$ | −0.01 | −0.16 | −0.03 | −0.37 | −0.01 | −0.06 | −0.09 | −0.26 | −0.25 | −0.47 |
|  | 0.17 | −0.04 | 0.46 | −0.15 | 0.60 | 0.68 | 0.73 | 0.15 | 0.12 | 0.27 |
|  | −0.01 | −0.10 | −0.02 | −0.29 | −0.02 | 0.19 | −0.00 | −0.52 | 0.38 | −0.21 |
|  | −0.03 | −0.33 | −0.05 | −0.41 | −0.01 | −0.07 | 0.03 | −0.37 | −0.28 | 0.10 |
|  | 0.40 | −0.10 | 0.57 | −0.19 | −0.10 | −0.00 | −0.22 | 0.10 | 0.02 | 0.02 |
|  | −0.02 | −0.21 | −0.02 | −0.35 | −0.01 | −0.14 | 0.05 | −0.52 | 0.35 | 0.20 |
|  | −0.04 | −0.48 | −0.02 | −0.01 | 0.00 | 0.08 | −0.01 | −0.19 | 0.13 | 0.33 |
|  | 0.56 | −0.12 | 0.02 | −0.00 | −0.47 | −0.30 | 0.08 | 0.14 | 0.01 | 0.01 |
|  | −0.04 | −0.30 | −0.01 | −0.01 | 0.01 | −0.13 | −0.35 | 0.03 | 0.13 | 0.16 |
|  | −0.07 | −0.56 | 0.04 | 0.47 | 0.00 | 0.21 | −0.29 | −0.05 | 0.54 | −0.48 |
|  | 0.66 | −0.14 | −0.60 | 0.18 | 0.39 | 0.07 | 0.15 | 0.01 | −0.16 | −0.07 |
|  | −0.04 | −0.34 | −0.01 | 0.40 | 0.01 | 0.31 | −0.01 | −0.09 | 0.07 | −0.12 |
| $\Phi(i)$ | 0.00 | 0.00 | 0.00 | 0.00 | 0.00 | 0.00 | 0.00 | 0.00 | 0.00 | 0.00 |
|  | 0.04 | −0.00 | 0.15 | −0.07 | −0.38 | 0.34 | 0.04 | 0.22 | −0.01 | 0.14 |
|  | −0.00 | 0.00 | −0.00 | −0.00 | 0.02 | 0.02 | −0.01 | 0.11 | 0.05 | −0.01 |
|  | −0.00 | −0.00 | −0.00 | −0.01 | −0.00 | −0.00 | 0.02 | 0.03 | 0.07 | 0.05 |
|  | 0.10 | −0.01 | 0.19 | −0.08 | 0.04 | −0.04 | −0.33 | 0.20 | 0.15 | −0.04 |
|  | −0.00 | −0.00 | −0.00 | −0.02 | −0.00 | −0.02 | 0.04 | 0.04 | 0.11 | −0.02 |
|  | −0.00 | 0.00 | −0.00 | 0.00 | −0.01 | 0.00 | 0.01 | 0.02 | 0.03 | 0.06 |
|  | 0.14 | −0.01 | 0.00 | 0.01 | 0.26 | −0.29 | 0.21 | 0.22 | −0.12 | −0.26 |
|  | −0.00 | −0.00 | −0.00 | 0.00 | −0.01 | −0.03 | 0.02 | 0.00 | 0.00 | 0.02 |
|  | −0.01 | 0.00 | −0.00 | 0.01 | 0.02 | 0.00 | −0.03 | 0.01 | −0.14 | −0.08 |
|  | 0.16 | −0.01 | −0.21 | 0.08 | −0.20 | 0.08 | 0.11 | 0.01 | −0.39 | 0.36 |
|  | −0.00 | 0.00 | −0.01 | 0.02 | 0.00 | 0.04 | −0.05 | 0.01 | −0.01 | −0.03 |

$$D_{KL}(P||Q) = \int_{-\infty}^{\infty} p(z) \log \frac{p(z)}{q(z)} dz \quad (11)$$

where $p(z)$ and $q(z)$ are the probability density functions of $P$ and $Q$. Some useful properties are that $D_{KL}(P||Q) \geq 0$, with equality if and only if $P = Q$, and that it is not symmetric, i.e., $D_{KL}(P||Q) \neq D_{KL}(Q||P)$.

In this work, the prior distribution $q(z)$ is assumed to be of zero mean value and unit covariance matrix, implying independence and equal variance for all degrees-of-freedom. This ensures that residuals are not correlated and have uniform variance, which is ideal when the model treats all degrees-of-freedom equally. An alternative investigation is shown in Sec. 6. The posterior distribution $p(z)$ has mean value equal to the result at each timestep, namely, $z[k] = x[k] - y[k]$ with covariance to be the covariance of the residual vector $z[k]$. By minimizing the Kullback–Leibler divergence with respect to various damping models after implementing an operation modal analysis, an information-theory-based model is created, which captures alert events and alert duration more reliably.

## 3 Application to the Multi-Axis Simulation Table at the University of Bath

The proposed methodology is applied to the multi-axis simulation table at the University of Bath; a state-of-the-art vibration test facility located within the Center for Power Transmission and Motion Control. The system is seen in Fig. 1 allowing for multidegree-of-freedom vibration testing for various applications, including earthquake simulation, vibration isolation assessment, and system verification under complex excitation conditions [60,61].

The table is a six-degree-of-freedom system, allowing motion in Cartesian coordinates along with rotational degrees-of-freedom (roll, pitch, and yaw). It is capable of handling payloads up to 450 kg, with test frequencies reaching up to 50 Hz. The platform measures $1.5 \times 1.7$ m and includes a grid of threaded holes for secure specimen mounting. The system is controlled via an Instron 8800 controller and offers cyclic, random, and user-defined input signal capabilities. Data acquisition is performed at a rate of 2.5 kHz, capturing six position signals and six acceleration signals from sensors mounted either on the table or the test specimen. During the test, data were collected for vertical movement of the table in an impact-type base excitation. The reference acceleration is then used within the model updating framework to identify the dynamic properties of the system as well as the early warning capabilities of the examined methodology.

To create an accurate model, the identified damping matrix for an assumed single degree-of-freedom vertical behavior is scaled with five different scalar values ranging from 10% to 500% of what the standard frequency domain decomposition method provides (scaling factor =[0.1 1 2 4 5], where each scaling factor value refers to model 1, 2, 3, 4, and 5, respectively, as seen in Fig. 2). This process takes place after the matching scaling of the response with respect to the maximum acceleration measurements and the maximum modeled acceleration.

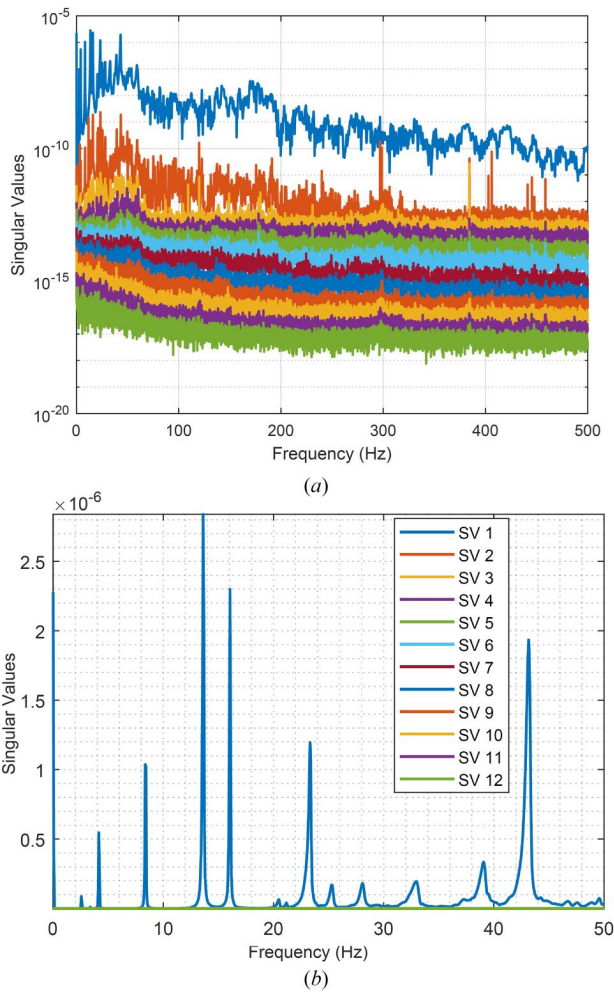

Fig. 4 Singular value decomposition values for Sec. 4 application (a) derived from the frequency domain decomposition technique [66], highlighting also the 10 highest singular values with a specific focus on singular value 1 (b)

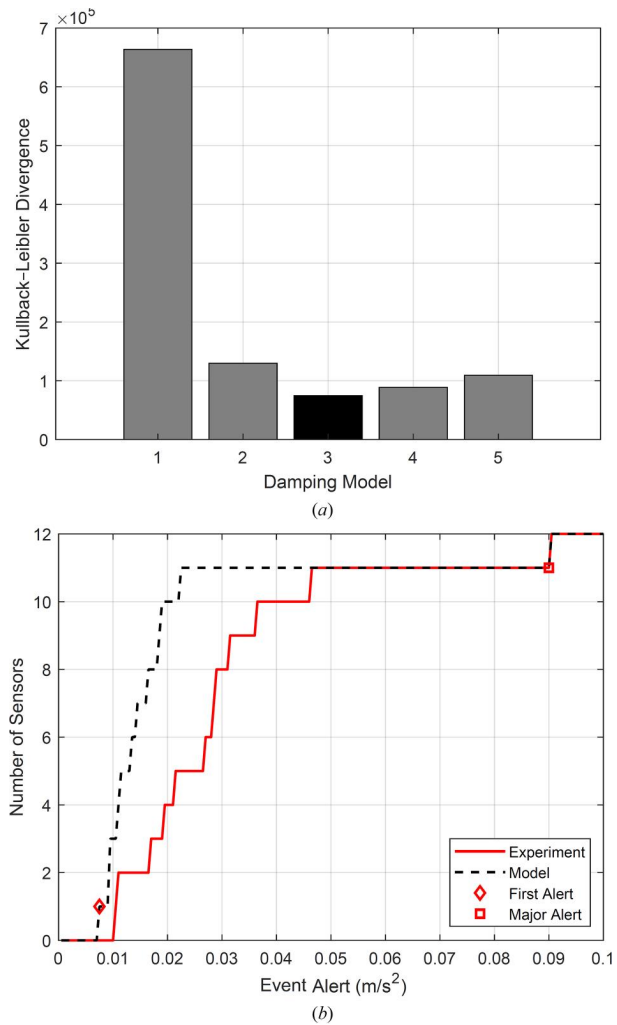

Fig. 5 Section 4 Kullback–Leibler divergence (a), and distribution of alerts (b). Plot (b) shows the amplitude of an event so that a sensor is triggered, i.e., for an event of less than 0.01 m/s² only one sensor is triggered, while for an event of 0.09 m/s² all sensors are triggered.

Figure 2 compares the time-history performance of the model to experimental data for all damping models to replicate experimental observations under different damping scenarios. Ideally, the model and the experimental responses should match. The input of impact-type base excitation is unknown, and output-only measurements are used.

Figure 3 shows the Kullback–Leibler divergence for various damping models. Figure 3 also compares the event duration between the experimental data and the model for various alert thresholds from 0 to 40 m/s$^2$ acceleration, i.e., the total time for which the signal exceeds the threshold. Ideally, the real duration of the alerts should match the model duration for the same alert. This comparison provides insights into how the model performs relative to the actual system behavior across different threshold settings.

The previous visualization pattern is also applied in the next section, where 12 sensors are used instead of just one. In that scenario, due to the larger number of sensors, simply examining time history plots is insufficient to evaluate the model's accuracy. Therefore, more summary plots are introduced. One comprehensive plot, for example, shows how many sensors are triggered (i.e., exceed the alert threshold) at each threshold level. Again, the model should aim to match the experimental results across all these metrics.

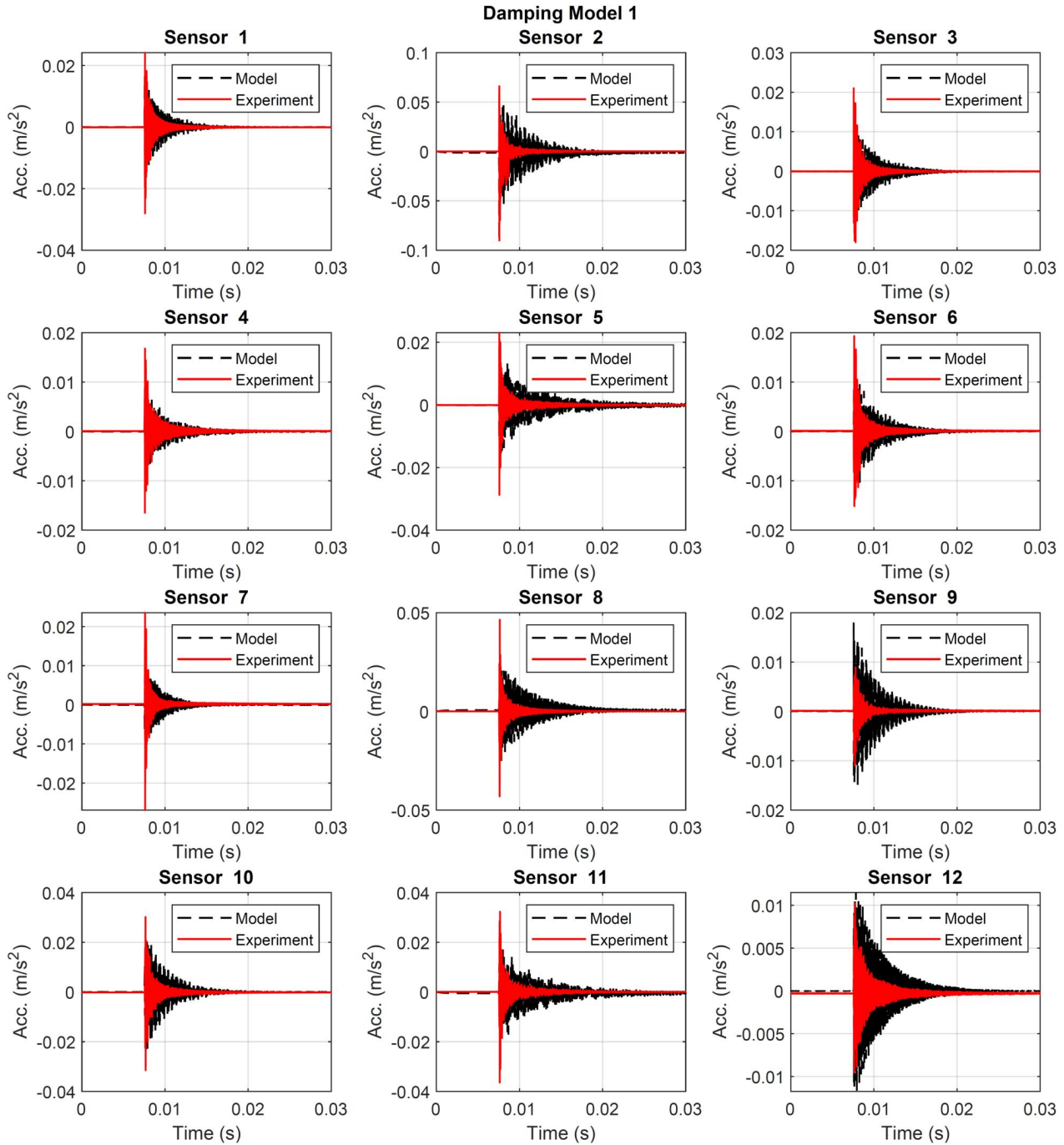

**Fig. 6  Time-history performance of the Sec. 4 system for the damping model 1**

## 4 Application to the International Association for Structural Control–American Society of Civil Engineers Benchmark Structural Health Monitoring Problem

The proposed methodology is also applied to the second phase of experiments conducted by the IASC–ASCE Structural Health Monitoring Task Group at the University of British Columbia [62–65]. The study focuses on applying structural health monitoring techniques to data collected from a four-story, two-bay by two-bay steel-frame structure. The structure, measuring $2.5 \times 2.5$ m in plan and 3.6 m tall, is mounted on a concrete slab outside the testing laboratory. To give a realistic mass distribution, floor slabs are placed in each bay per floor, with off-center masses on each floor [62]. The experimental setup included three types of excitation: electrodynamic shaker, impact hammer, and ambient vibration. Accelerometers strategically placed across the structure facilitated the measurement of structural responses. Fifteen accelerometers were positioned throughout the frame and the base to capture the responses of the test structure. The placement included sensors for measuring northsouth and eastwest motion. The excitation and impact hammer tests employed a Dynatron 5803A (Dytran Instruments, Chatsworth, CA) 12 lbf Impulse Hammer. This hammer, equipped with a force transducer, recorded measurements during tests involving 3–5 hits. Impact locations were chosen on the south and east faces of the first floor in the southeast corner. A 16-channel DASYLAB acquisition system recorded structural responses, with sampling rates of 250 Hz for shaker and ambient tests, and 1000 Hz for hammer tests. Anti-aliasing filters were applied selectively, and the data acquisition system commenced before the first impact, recording a series of hits within each test.

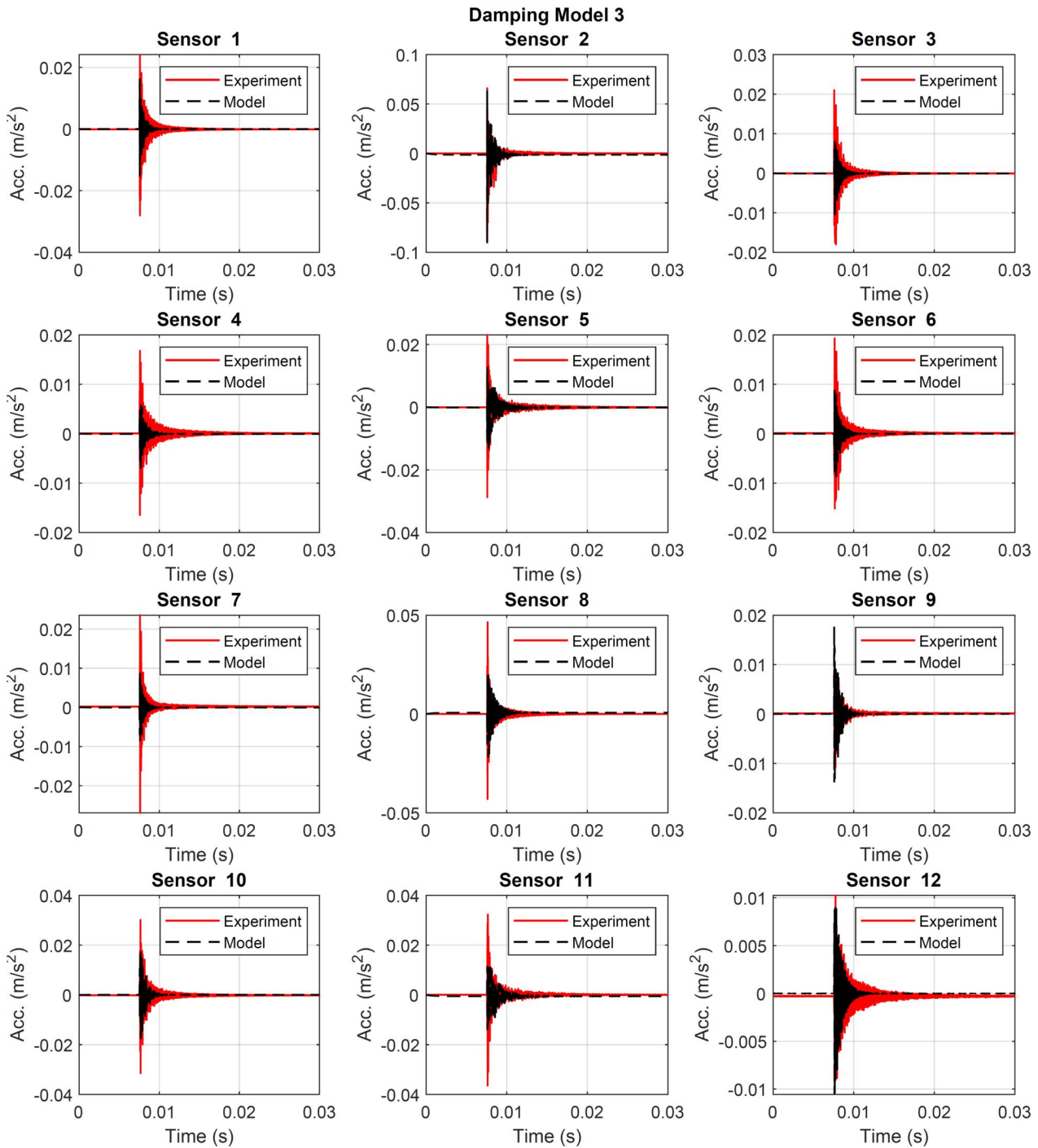

Fig. 7 Time-history performance of the Sec. 4 system for the damping model 3

Table 1 provides a summary of the identified frequencies, damping ratios, and the real and imaginary parts of the mode shapes for the case 9 damage configuration [62]. The data used are from the horizontal hammer impact on the south face. Importantly, the methodology is applied to create a model for the structure, while the damage detection capabilities are shown separately in Sec. 5.

Figure 4 shows the singular value decomposition values derived from the frequency domain decomposition technique [66]. The figure highlights the 10 highest singular values, with a specific focus on Singular Value 1, which is identified as the dominant singular value in this analysis.

To create an accurate model, the identified damping matrix is scaled with five different scalar values ranging from 10% to 150% of what the standard frequency domain decomposition method provides (scaling factor =[0.1 0.3 0.7 1 1.5], where each value refers to model 1, 2, 3, 4, and 5, respectively, as seen in Fig. 5). This process takes place after the matching scaling of the response with respect to the maximum acceleration measurements and the maximum modeled acceleration. Figure 5 provides the

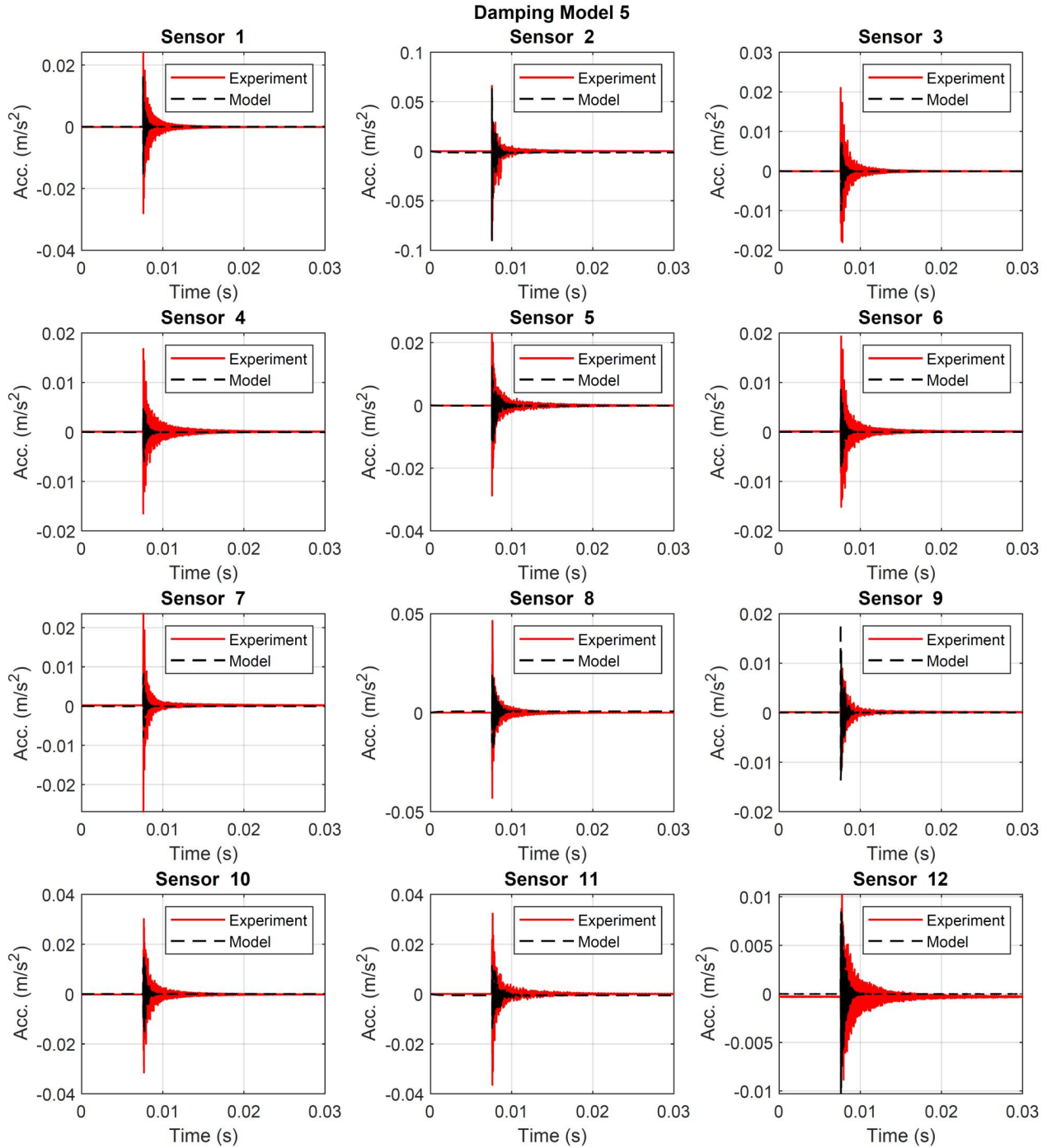

Fig. 8  Time-history performance of the Sec. 4 system for the damping model 5

051009-8 / Vol. 148, SEPTEMBER 2026Transactions of the ASME

Kullback–Leibler divergence for various damping models, and the distribution of alerts generated at each alert threshold for the optimal model, the one with the least Kullback–Leibler divergence. The last plot is particularly useful for understanding the number of sensors activated during different alert events, namely, it demonstrates how many sensors exceed the acceptable vibration level each time. It is seen that the first alert is triggered when at least one sensor exceeds the danger threshold value, while the major alert event is at the moment where all sensors are triggered, namely, all sensors exceed this threshold.

Figure 6, then, compares the time-history performance of the model to experimental data for the damping model 1, namely, the first candidate scaled damping matrix. The response is shown for a horizontal hammer-type excitation of the case 9 damage configuration, as described earlier. Ideally, the model and the experimental responses should match. Importantly, comparisons for the undamaged configuration are shown later in Sec. 5, and here only the development of the model is of interest. Similarly, Figs. 7 and 8 show the time-history response comparisons for damping models 3

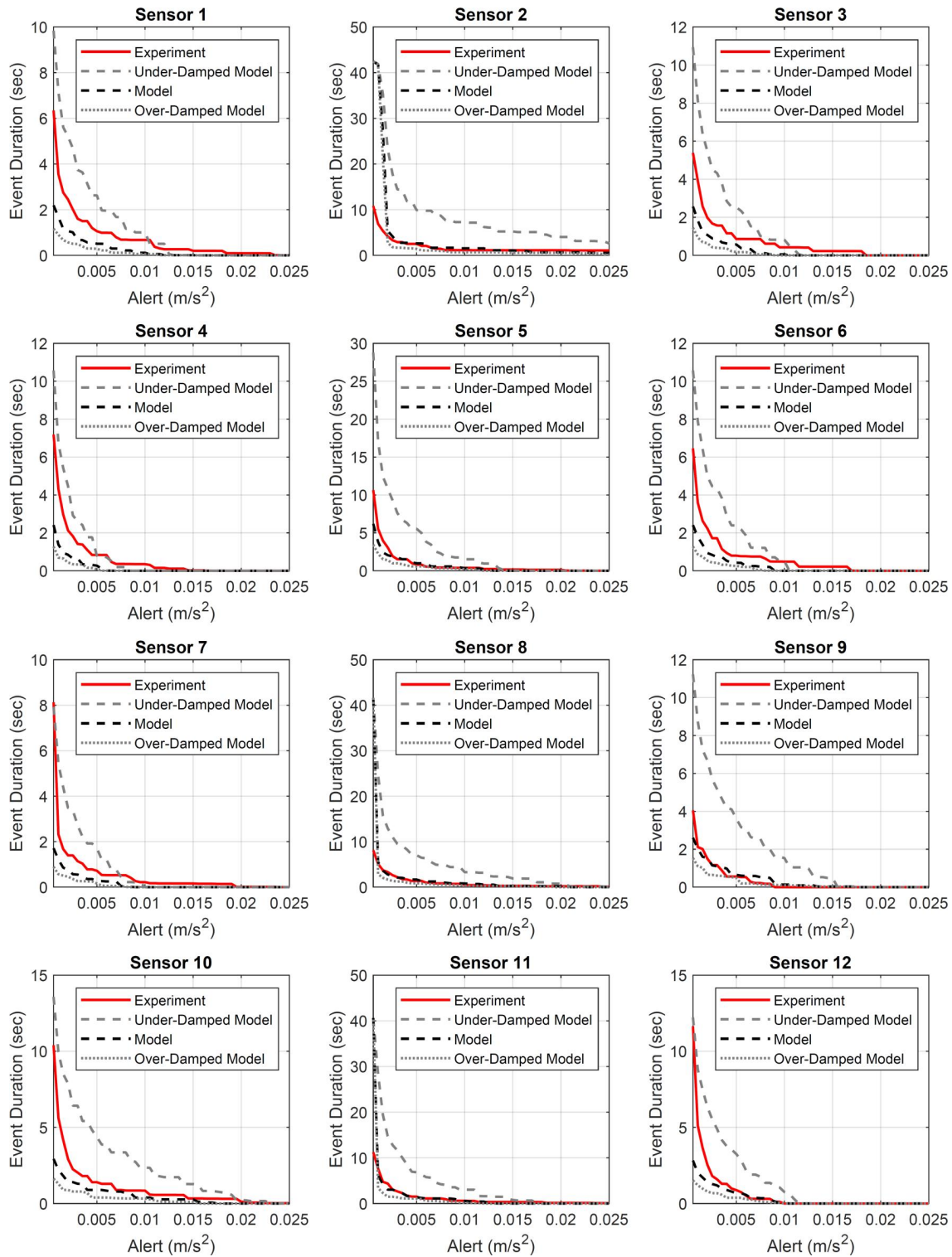

Fig. 9 Event duration for the Sec. 4 system for various alert thresholds

and 5, respectively, illustrating the model's ability to replicate experimental observations under different damping scenarios.

A detailed analysis is then performed to compute the duration where the measured and modeled accelerations exceed certain thresholds; this results in a different result plotting visualization. It is done for all sensors and is used to evaluate the model's performance for different limits. Figure 9 compares the event duration between the experimental data and the model for various alert thresholds.

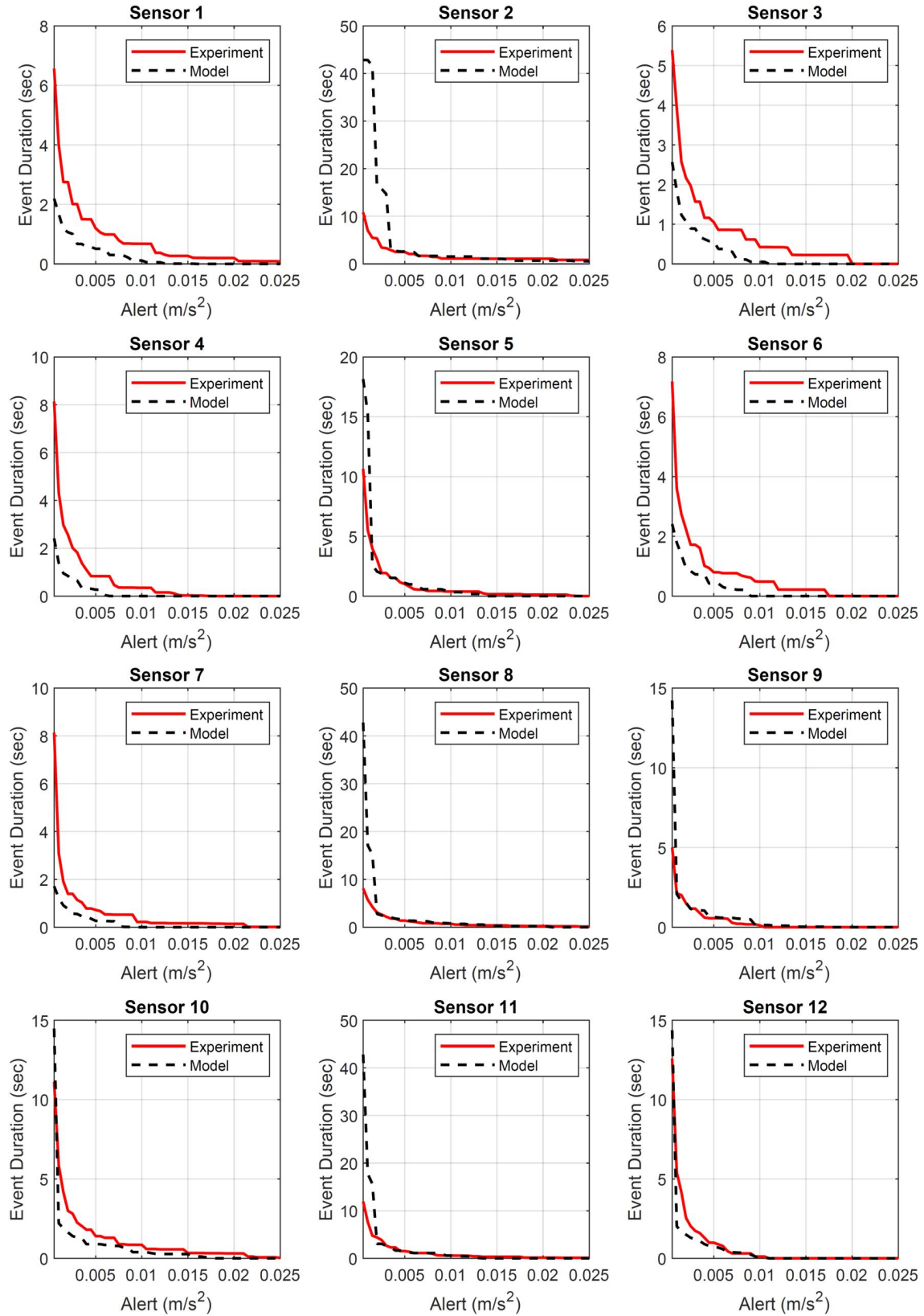

Fig. 10 Event duration for the Sec. 4 system for a hammer-type loading different than the one used to identify the structure (same damage case 9 configuration [62])

This comparison provides insights into how the model performs relative to the actual structural behavior across different threshold settings. Ideally, the real duration of the alerts should match with the model duration for the same alert. Importantly, it shows the optimal model (labelled as model), as well as the under-damped model 1 and the over-damped model 5 as seen in Fig. 5(*a*).

Finally, Figs. 10 and 11 extend this comparison to include different impact load scenarios, of the same damage case 9 configuration [62] and of a similar load amplitude. These figures show the event duration for different loads, providing a broader understanding of the model's response to varying dynamic excitations.

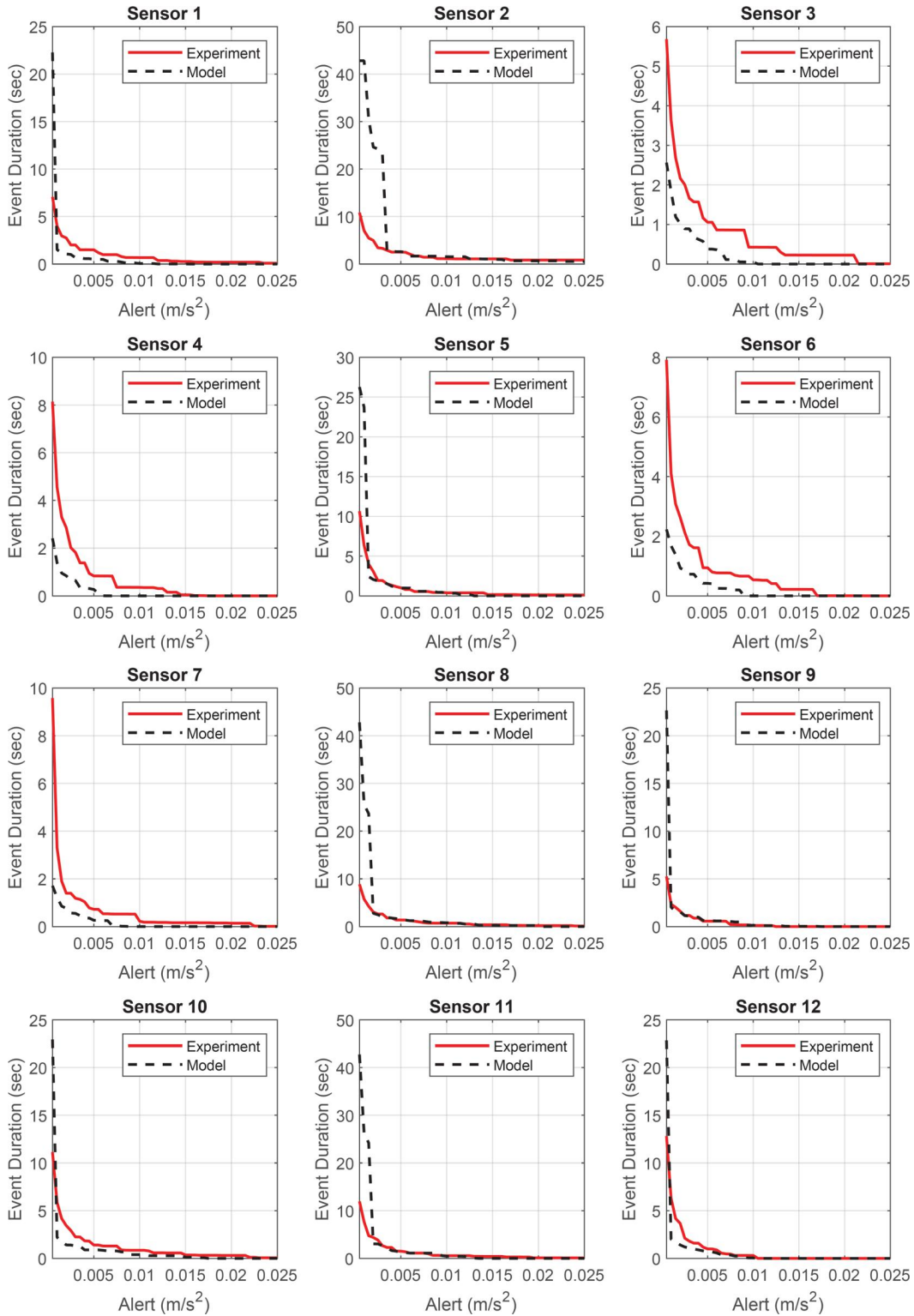

Fig. 11 Event duration for the Sec. 4 system for a second hammer-type loading different than the one used to identify the structure (same damage case 9 [62] configuration)

## 5 Damage and Anomaly Detection

The examinations so far showed the potential of the method when the structural model remains the same. However, this assumption may not be true if the structure changes, due to some damage, for instance, or any other modification of the structure [67,68]. This section, therefore, examines the extrapolation capabilities of the approach, since only a trained model was considered so far.

To explore this, consider the examined dynamic system. Compared to the previous structure of Sec. 4, diagonal elements have been added to create a stiffer structure [62]. This simulates a different damage level scenario (cases 7 and 1 [62]).

Figure 12 shows the overall performance of each sensor for a minor structural change (case 7). The system performs reasonably well for small structural changes.

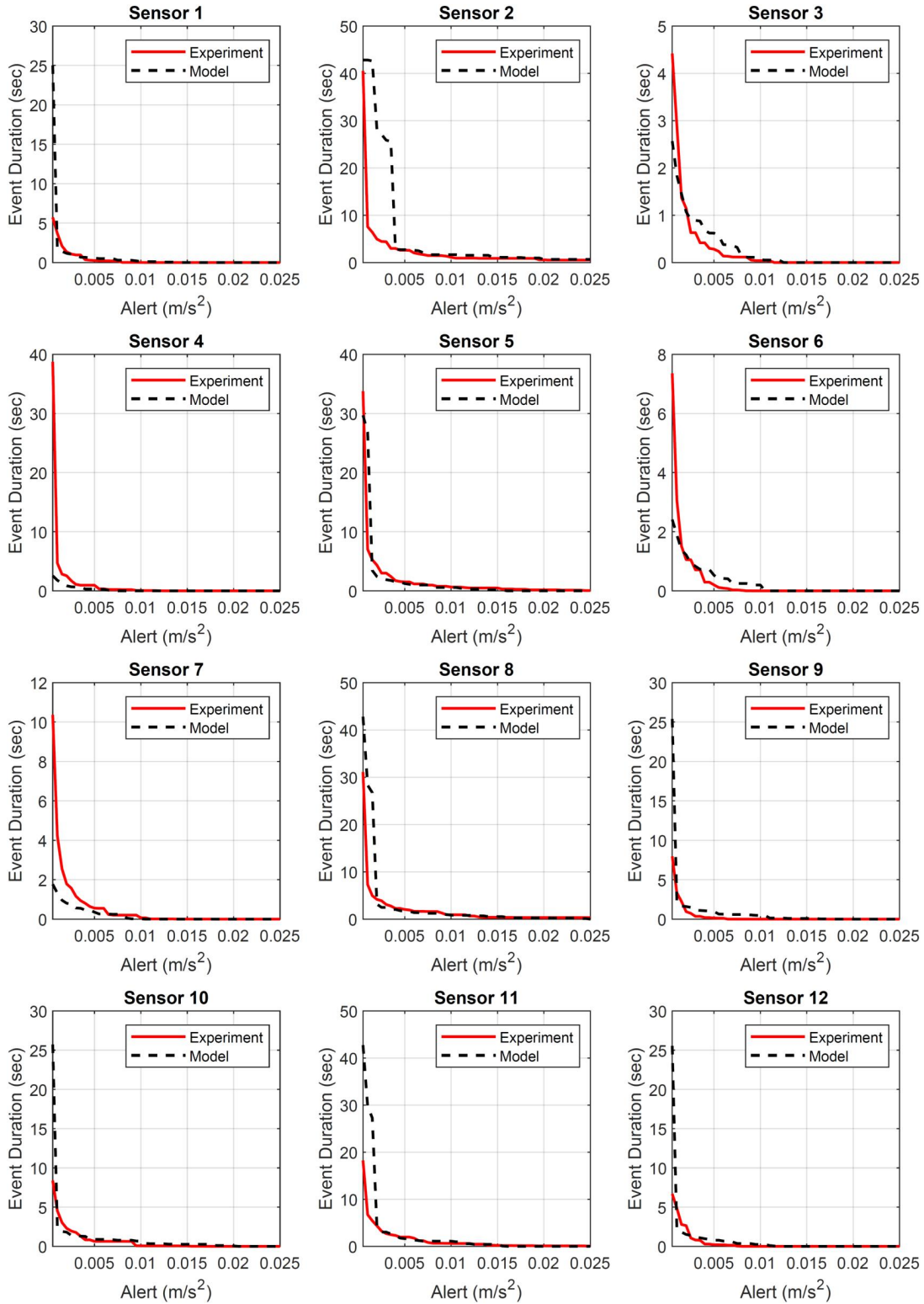

**Fig. 12** Section 5 model for a hammer-type loading different than the one used to identify the structure and for a different damage level of the structure (case 7 [62])

Figure 13, however, provides results for a significant structural change (case 1). The performance here decreases, where more substantial errors are observed for major events due to the extent of the structural modification, demonstrating the system's sensitivity to large changes in the structure.

Finally, Fig. 14 illustrates the overall performance of each sensor for a minor structural change (case 9) under a shaker load, so significantly different from the hammer excitation used to identify the model. Notably, the results remain satisfactory after a new model identification following the same process as in Sec. 4 from scratch, which highlights the need for periodic model recalibration in practical applications.

As a result, the approach is not capable of extended extrapolation to predict alerts with forcing or structural conditions outside of the training dataset with sufficient accuracy. When employed on a real engineering system where the system may change, one must have some prior knowledge about the expected forcing patterns in order to generate comprehensive training datasets and retrain the model for future good prediction. It follows, as a future recommendation, that the method to be combined with respect to online [69] or Bayesian methods [70–73] to incorporate prior knowledge. This observation, though, provides an indication of structural change (damage or anomaly in the conditions) when the deviation starts to become large compared to the initial calibrated model. Moreover, this modeling approach allows for faster and more intelligent early warnings than output thresholding alone. Because the model encapsulates the dynamic properties of the structure, it can identify deviations from expected behavior before thresholds are breached, offering predictive warning rather than reactive detection.

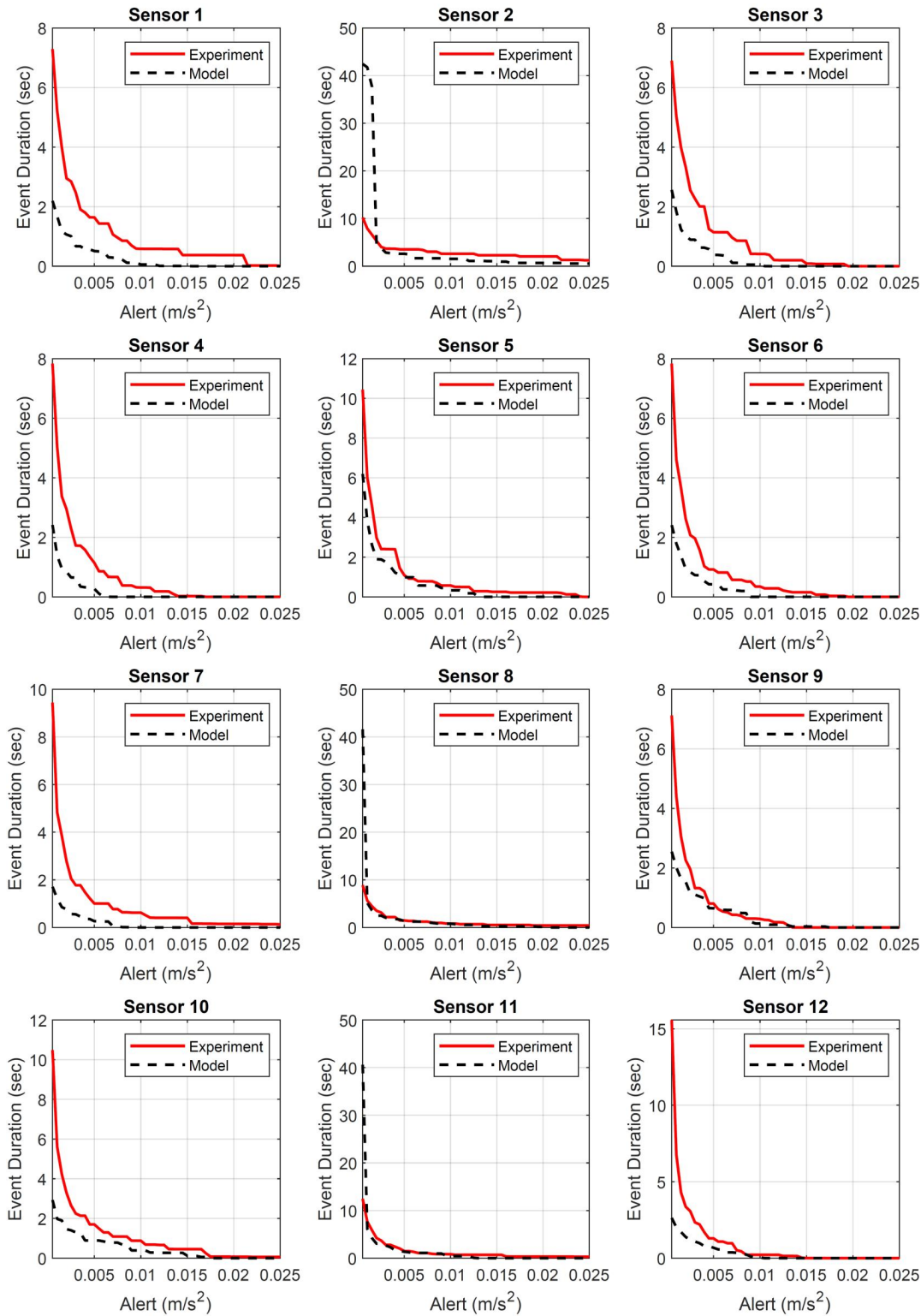

**Fig. 13 Section 5 model for a hammer-type loading different than the one used to identify the structure and for a different damage level of the structure (case 1 [62])**

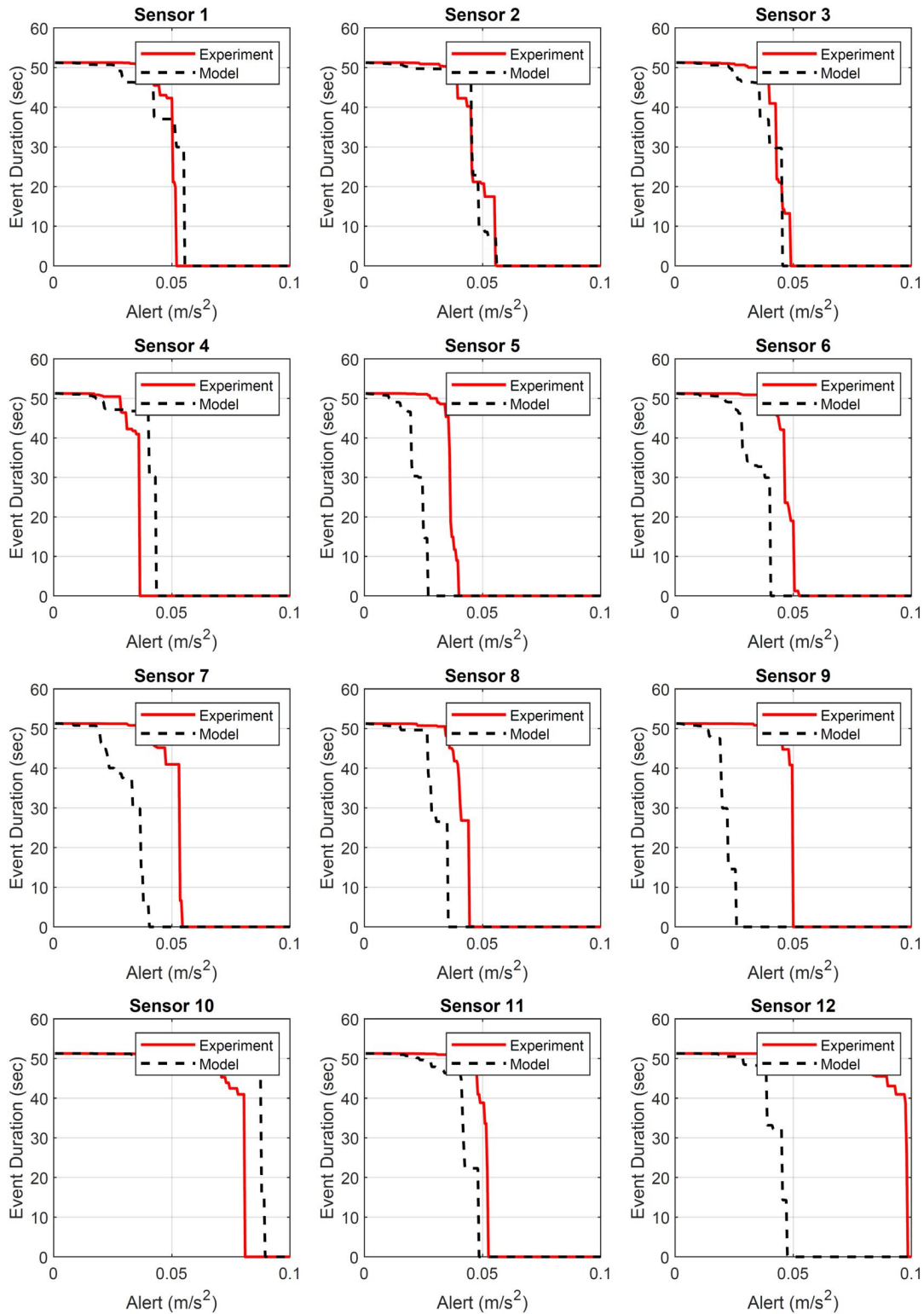

Fig. 14 Section 5 model for a shaker-type loading different than the one used to identify the structure and for a different damage level of the structure (case 9 [62])

## 6 Discussion

The presented work provided an information-entropy method to select the optimal dynamic model updating approach [74–79], without any input information [80,81], for accurate early warning system implementation. To this end, the robustness of the method was tested using both benchmark data and new experimental data.

Importantly, the final updated model has the mathematical form of Eq. (1) scaled to the unknown input magnitude. The methodology is based on the standard second-order dynamic system representation, which forms the backbone of structural dynamics. This was chosen deliberately to maintain generality and physical interpretability across structural systems, particularly under operational and output-only conditions [82,83]. The damping matrix is uniformly scaled for the entire structure without assigning separate damping values for different components. Although it seems that localized damping scaling is ignored, it is applied in this way to avoid the need for prior knowledge requirements for a structural system, or of sensor configuration. In this way, a system-level global identification is examined. In future implementations, this could be extended to an online or adaptive update scheme as the damping itself can also change depending on the structural modification, loading conditions, or even environmental variability [84,85].

Relating to the rationale behind the method, the signal energy of the residuals provides a direct and physically meaningful measure of the overall mismatch between the measured and modeled responses. To further justify this choice, Fig. 15 presents a comparison with two widely used alternatives: Rényi entropy and Jensen–Shannon divergence. Unlike Kullback–Leibler divergence, which exhibits a clear and unique minimum that consistently identifies the optimal damping model, both Rényi and Jensen–Shannon measures produce a behavior without distinct optimal values, indicating insufficient discriminatory capability for damping estimation. Specifically, the Kullback–Leibler divergence consistently shows a distinct minimum at a mid-range damping value, indicating the point at which the modeled response best matches the measured data. In contrast, the Rényi and Jensen–Shannon divergences exhibit mainly monotonic patterns as damping increases. These metrics are more smoothed, symmetric, and bounded, making them less responsive to subtle variations in distribution shape and more influenced by the overall energy trend, which decreases monotonically with increasing damping in physical systems. Regarding using other types of information entropy or other intelligent measures [86], future research investigation is required to judge which model performs better.

A concern is related to the computational cost when all sensor data are considered. The computational cost of this approach though is acceptable. This is attributed to three main reasons: the one-dimensional nature of the data, the simultaneous processing of the operation modal analysis algorithms, and the analytical solution of the Kullback–Leibler divergence when the Gaussian assumption holds. Table 2 shows the computational time for both experimental investigations. These results demonstrate that the full model-selection pipeline, including operational modal analysis, residual computation, and information-theoretic evaluation, executes within a near-real-time environment for a common modern computer with an Intel Core Ultra 7 155 U processor and 16GB of RAM. Importantly, once the optimal model is identified, the real-time stage requires only the residual evaluation and Kullback–Leibler-based alert logic, which reduces the computational cost. The University of Bath multi-axis simulation table (MAST) experiment was intentionally simplified to a single-degree-of-freedom vertical model to allow controlled validation of the methodology under a setting where the input direction is known, and cross-axis coupling is minimal. However, the method is not limited to SDOF (single degree of freedom) systems. Its applicability to multi-degree-of-freedom (MDOF) dynamic behavior is demonstrated through the IASC–ASCE benchmark structure, which includes 12 sensors and full 3D vibration coupling, providing a realistic MDOF environment with extensive mode interaction.

**Table 2  Computational time for the MAST platform and the IASC–ASCE benchmark problem**

| MAST single-DOF experiment | Run time (s) |
| --- | --- |
| Full implementation (all steps) | 2.2233 |
| Before information-theory processing | 0.032821 |
| If damping is known (identified model) | 0.70668 |
| IASC-ASCE multi-DOF Benchmark | Run time (s) |
| Full implementation (all steps) | 21.0896 |
| Before training starts | 0.054129 |
| If damping is known (identified model) | 2.8839 |
| If the model is known (new tests only) | 5.7353 |

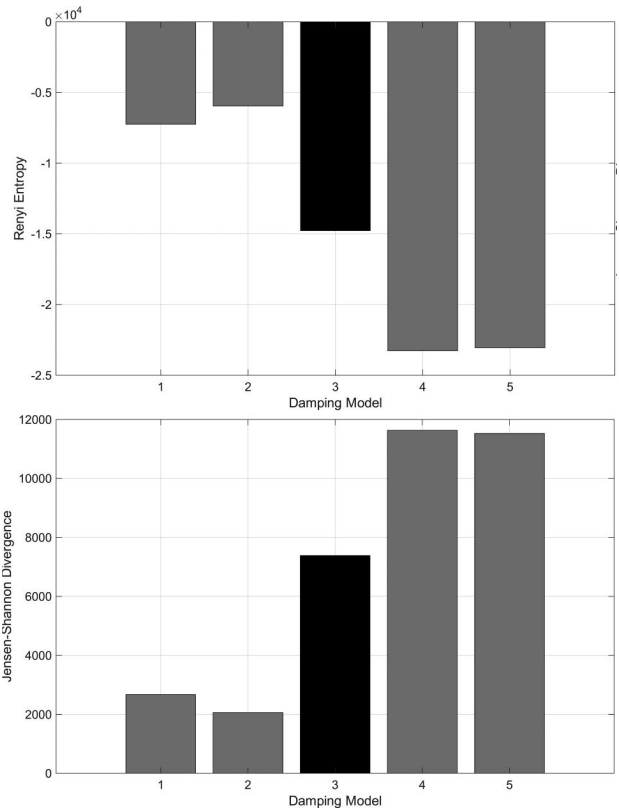

**Fig. 15  The Renyi entropy and the Jensen–Shannon divergence for the Sec. 3 system**

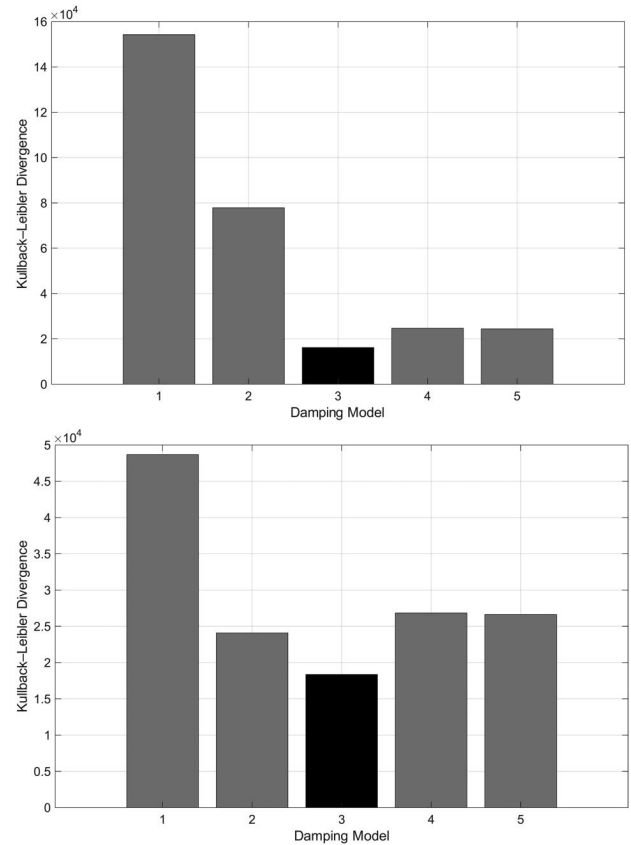

**Fig. 16  Prior distribution investigation for the Sec. 3 system (covariance magnitude 0.5 and 2, respectively)**

**Table 3 Comparison of natural frequencies between reference and a slightly damaged state**

| Mode | Reference frequency (Hz) | Damaged frequency (Hz) |
| --- | --- | --- |
| 1 | 2.564 | 2.525 |
| 2 | 4.150 | 4.134 |
| 3 | 8.362 | 8.562 |
| 4 | 13.611 | 13.916 |
| 5 | 16.052 | 16.174 |
| 6 | 23.315 | 23.174 |
| 7 | 25.269 | 25.606 |
| 8 | 32.959 | 32.973 |
| 9 | 39.063 | 39.452 |
| 10 | 43.152 | 43.457 |

The sensitivity of the result to the assumed prior distribution is also of interest. Figure 16 shows a comparative test in which the prior distribution was altered to have (i) a zero-mean value with 0.5 magnitude for the covariance, and (ii) an inflated covariance of magnitude 2. The resulting Kullback–Leibler divergence behavior remains nearly unchanged, and the algorithm reliably selects the same optimal damping model in all cases. This confirms that the method's performance is robust to reasonable variations in the assumed prior distribution, and that the latter primarily functions as a reference baseline rather than a parameter that materially influences model selection.

Related to detecting structural modification even in cases where modal parameters exhibit minimal or nondiagnostic changes, such as in cases 7 and 9 [62], Table 3 is provided, which shows that the frequency changes between the reference and slightly modified configurations remain extremely small, especially in the low-frequency rigid-mode range ($<5$ Hz). Despite these negligible modal shifts, the proposed approach successfully identifies the presence of damage through statistically significant deviations in the event-duration metrics. This highlights a key advantage of the method over classical modal-based approaches.

Uncertainty may also be quantified using the standard deviation of the residuals, which directly reflects the variability of the prediction error. The computed values in Table 4 demonstrate that the MAST SDOF system exhibits a very low residual spread, whereas the more complex MDOF benchmark, containing multiple coupled modes and significantly more sensors, shows a correspondingly larger but still well-bounded residual spread. While full Bayesian uncertainty quantification [87–91] is an important direction for future work, the added residual-spread analysis offers a computationally lightweight uncertainty indicator aligned with near real-time monitoring requirements. This reduces the risk for the posterior outcome to be attached to algorithmic parameters, for instance, in Kalman filtering process and observation matrices [92–95].

A final concern is related to the integration of the proposed forecasting framework within a broader model updating paradigm. The digital twin concept is a relatively new topic [96]. It consists of four essential components: (a) the physical entity and the data collection from it, (b) the development of a virtual representation or simulation of this entity, (c) the continuous interaction and synchronization between the physical and virtual systems, and (d) the application and predictive capabilities enabled by this coupling. In this context, the present work aims to establish a digital twin model by learning and updating the system's dynamic behavior.

## 7 Conclusions

The information theory-based model updating capabilities were examined here. The current output-only operational modal identification methods, such as frequency domain decomposition or stochastic subspace identification, suffer from poor estimation of the damping behavior of a system.

Overall, the proposed approach allowed for system identification with:

(1) Simultaneous identification of multiple models.
(2) Statistics-based comparison of damping performance with a full duration modeling of vibration alerts.
(3) Integration of all structural sensors into one index with automatic selection of the identified model with the most plausible results.
(4) No need for data filtering and with a minimum extra computational cost to the standard operation modal analysis procedure.
(5) Near real-time application when the system identification has been completed.

Importantly, the approach is shown to select the optimal model, which accurately captures the correct alert duration, providing a powerful tool for early warning systems.

## Acknowledgment

The authors gratefully acknowledge Jens Roesner, Technical Manager of the Centre for Power Transmission and Motion Control, for his assistance in acquiring measurements from the University of Bath shaking table, and the Structural Health Monitoring Task Group for providing the IASC–ASCE benchmark structural health monitoring problem data.

**Table 4 Uncertainty quantification based on the standard deviation of residuals from the optimal damping model**

| System | Residual standard deviation |
| --- | --- |
| MAST (SDOF) | 0.001573 |
| IASC-ASCE benchmark (MDOF) | 0.081494 |